\documentstyle[12pt]{article}
\begin{document}
\topmargin=-.5in
\oddsidemargin=.1in
\evensidemargin=.1in
\vsize=23.5cm
\hsize=16cm
\textheight=23.0cm
\textwidth=16cm
\baselineskip=24pt
\thispagestyle{empty}

\hfill{ITP-SB-98-32}\\
\smallskip
\hfill{ May 1998}

\vspace{1.0in}

\centerline{\Large \bf Fractional Quantum Hall Layer As } 

\centerline{\Large \bf A Magical Electromagnetic Medium}

\vspace{.85in}

{\baselineskip=16pt
\centerline{\large Alfred Scharff Goldhaber}
\bigskip
\centerline{\it Institute for Theoretical Physics}
\centerline{\it State University of New York}
\centerline{\it Stony Brook, NY 11794-3840}}

\vspace{.5in}

\centerline{\Large Abstract}

\vspace{.5in}

 At a surface between electromagnetic media
the Maxwell equations are consistent with either the usual boundary conditions,
or exactly one alternative: continuity of  
${\bf E_\perp}$,
 ${\bf H_\perp}$, ${\bf D_\parallel}$, ${\bf B_\parallel}$.
These alternative, classically inexplicable conditions 
applied to the top and bottom surfaces of an FQH layer capture 
exactly its unique
low-frequency properties.

\newpage

``Magic'' in science means not illusion or trickery, but rather phenomena so
surprising and counterintuitive that even when explained
they inspire awe.  An example in classical physics is the
experiment of holding a tennis ball just above a basketball at chest height over
a hard floor, releasing the balls simultaneously, and seeing the tennis ball
bounce up to hit even a very high ceiling!  That this follows directly from
conservation of energy and momentum makes it no less amazing.  Quantum physics
contains much that is magic in this sense, and both the original \cite {K} and 
the  fractional \cite{TS} quantum Hall  effects are striking illustrations.

The Hall effect is a steady current traveling perpendicular to the plane
defined by crossed electric (${\bf E}$) and magnetic (${\bf B}$) fields.  The
current may be described as due to an array of charges all traveling in the same
direction with speed
$v={\bf |E|/|B|}$.  In classical physics the number of these charges and
hence the magnitude of the current can be arbitrary.  In the quantum Hall
effect, for a given sample the magnetic field may be varied over a substantial
range without changing the effective number of charges $\nu$ per quantum of
magnetic flux contributing to the Hall current.  The value of $\nu$ is an
integer for the original quantum effect, and a rational fraction for the
fractional [FQHE] version.

FQHE is remarkable for many reasons,
not least that in a very short time it found its `standard model' in the
composite-fermion picture \cite{J}, which unifies Laughlin's original description
of simple Hall fractions $\nu = \frac{1}{2p+1}$ \cite{Laugh} with a host of other
observed phenomena in the fractional Hall domain, as well as earlier
understanding \cite {Laugh0} of the integer quantum Hall effect. 
Perhaps because progress in microscopic theory was so rapid, a
familiar stage from previous studies of macroscopic systems -- phenomenological
description in terms of an electromagnetic medium -- appears to have been
skipped.  Another reason for this omission may be that Hall samples
characteristically have macroscopic (${\cal O}({\rm cm})$) dimensions in the
directions perpendicular to the magnetic field, but a thickness of only
${\cal O}(500 { \rm \AA})$ in the parallel direction, so that a
macroscopic description conceivably might not even exist.     
 
 The
purpose of this work is to exhibit a unique,
consistent, yet unprecedented  option for characterizing an electromagnetic
medium which exactly captures the extraordinary properties of the fractional
quantum Hall layer, besides the Hall current itself.   This
macroscopic description indeed is magic in the sense mentioned above, because it
cannot be reproduced by any realistic classical model.  A more comprehensive
and detailed discussion
appears elsewhere \cite{ASG}, but the intention here is to make the basic idea
accessible to a wide audience. At the very least, this realization should be a
useful mnemonic device, but it also may help shape further insights into an
extraordinarily fascinating system.

Jain noted \cite{J,GJ} that FQHE can be described as combining a familiar
property, renormalization of local quasiparticle charge by polarization of the
medium, with an entirely novel property extending even beyond the fractional Hall
plateaux, renormalization of the perpendicular magnetic field and tangential
electric field inside the medium with respect to values in the external regions
immediately adjacent to the surfaces of the Hall layer.  
 These results for FQHE
may be summarized by the statements:

\noindent
1)  The Gauss-law charge of a quasiparticle, measured by its electric field far
away, is $e^\star$, where the ratio $e/e^\star = 2pn\pm 1$ is related to the
Hall fraction
$\nu = \frac{n}{2pn\pm 1}$.

\noindent
2)  Quasiparticles move in the presence of an  electromagnetic field 
as
if they carried electric charge
$e^\star$ -- or equally well as if they carried a charge $e$, but in
 fields
${\bf B}^\star _\perp=e^\star{\bf B}_\perp /e$ and  ${\bf E}^\star _\parallel
=e^\star {\bf E}_\parallel/e$.

Further, it is well accepted that 

\noindent
3)  If an electron were gently inserted into the surface of a Hall layer, it
would break up into $2pn\pm 1$ quasiparticles \cite{Su}.

A naive effort to describe 
the Hall layer as a conventional dielectric medium fails at once.  To take the
most elementary aspect, far away from a charge located in a (conventional) thin
dielectric layer there is no renormalization of the charge by the dielectric
constant of the medium, because the surface charge which together with the local
charge makes a total
$e$ is itself localized quite close to the particle.
By the same token, property 2) also does not hold.  Is there any viable
alternative?  

The usual
 surface boundary conditions are continuity of
${\bf D}_\perp , {\bf B}_\perp, {\bf E}_\parallel,{\bf H}_\parallel$.  There is
a well-established classical picture underlying these conditions, that
tiny electric and/or magnetic dipoles orient themselves under the influence
of applied fields, leading to effective surface charges 
and currents which, for example,
account for a discontinuity in the normal component of ${\bf E}$.
However, the Maxwell equations would remain consistent under simultaneous
interchange between the roles of ${\bf E}$ and ${\bf D}$, ${\bf B}$ and ${\bf
H},$ imposing continuity of
${\bf E}_\perp , {\bf
H}_\perp, {\bf D}_\parallel ,{\bf B}_\parallel$.  Note that just interchange
of one pair would not be consistent:  Such mixed
boundary conditions would violate the duality-rotation symmetry of
the Maxwell equations under which ${\bf E \rightarrow H}$ 
and ${\bf D \rightarrow B}$, while ${\bf H \rightarrow -E}$ and ${\bf B
\rightarrow -D}$.  
 Thus, there is only one mathematically consistent
 alternative to
consider as a possible physical description for FQHE.  Let us see if this sole
remaining
option works.

Assume
the dielectric constant $\epsilon$ and magnetic permeability $\mu$ are given by
$\epsilon = \mu ^{-1} = e/e^\star$.  Then continuity of $E_\perp$ at the top
and bottom surfaces assures that the total electric flux coming out of a charge
placed in the layer corresponds to a charge $e^\star$, not $e$, thus confirming
1).  Continuity of ${\bf H}_\perp$ yields a perpendicular magnetic field inside
the layer of magnitude $\mu B$, , and continuity of ${\bf D}_\parallel$ implies
a parallel electric field inside of magnitude $E/\epsilon$, confirming 2).
If an electron descends into the Hall layer, total local charge $e$ must be
conserved, but inside the layer each individual charge is $e^\star$.  

In a
conventional insulator, the electron would leave part of its charge on the
surface, but in
our
context, where distributing the residual charge in the surface of the
layer is not possible,   
the only way to achieve local implementation of the conservation law is by
generation of
$2pn\pm 1$ quasiparticles.  Thus, even fact 3) follows from this novel set of
constitutive relations.  Note that, although the treatment here is not
explicitly quantum-mechanical, local conservation of electric charge requires
quantization of $\epsilon$ at integer values, as otherwise conservation of
charge through breakup would not be possible \cite{Su}.  The fact that only
odd integers are allowed follows because at the core of each quasiparticle 
must be a fermion, and by a superselection rule an odd number of fermions may
not turn into an even number \cite{WWW}.

So far we have seen how the constitutive relations and surface boundary
conditions reproduce known results, but one might wonder if they can give any new
information.  One obvious question to address is whether the Hall layer exhibits
new `trapped' electromagnetic modes, which in turn might be detected by
scattering experiments.  
However, there can be no such modes because, even with our exotic boundary
conditions, trapping produced by critical internal reflection 
requires (relative) refractive index greater than unity inside, and in this case
we have exactly unity.  Furthermore, the fact that
the layer must be quite thin on the wavelength scale relevant to FQHE means that
even reflection of external waves must be strongly suppressed.  

Besides possible
optical modes, one might wonder if there could be a longitudinal mode, but any
such mode must have finite mass because of the incompressibility of the FQH
ground state in a specified perpendicular magnetic field.  Of course, for
compressible states, as near
$\nu=\frac{1}{2}$, there can be longitudinal modes, and consideration of these
has led to the suggestion that there may be  modifications of the
simple composite-fermion Fermi surface expected if these modes
were ignored \cite {HLR}.  Recent studies have focused on these longitudinal
`plasmon' modes to obtain remarkably detailed analytic results for both the
compressible regime and the incompressible regime (where the plasmon has an
effective mass) \cite {SM}. 

The dielectric response function of any medium should depend on
frequency and wavenumber.  Provided $\epsilon$ and $\mu$ approach unity
 as the frequency rises above some critical
value, the unique description proposed here should remain in
agreement with observation.  In particular, it should be possible to describe
the full frequency and wavenumber dependence of $\epsilon$, including a peak
in Raman response found for the $\nu=1/3$ state \cite {Pin}.  

Another aspect which deserves attention is the boundary conditions on the
edge surfaces of the FQH layer, which of course are quite narrow compared to
the top and bottom surfaces.  At an edge one must have (partially) the traditional
conditions.  This is immediately clear for ${\bf D_{\perp}}$ and ${\bf
B_{\perp}}$ and for the components of ${\bf H_{\parallel}}$ and ${\bf
E_{\parallel}}$ lying perpendicular to the Hall plane.  On the other hand,
consistency with conditions on the top and bottom surfaces also implies that the
components of
${\bf D_{\parallel}}$ and ${\bf B_{\parallel}}$ lying in the Hall plane 
should be continuous. 

At first sight these hybrid conditions might seem
unsatisfactory.  However, because the strong external magnetic field 
perpendicular to the layer determines
a special direction, these requirements are no more peculiar than the
familiar appearance of a complicated dielectric tensor in some anisotropic
medium.  
If anything, it is remarkable that in this case the electric and
magnetic  susceptibilities are just
scalars, with the anisotropy entirely described by the boundary conditions. 
One might ask whether the hybrid boundary conditions could lead
to interesting behavior near the edge of the sample.   Continuity of ${\bf
D_{\perp}}$ suggests that a single particle entering might proceed as a
quasiparticle into the medium, leaving the remainder of the charge to populate
edge states, and thus mimicking -- on the edge only -- the induced charge found
on the surfaces of a traditional insulator.  Whether this possibility gives any
useful insight about edge states remains to be seen.

We may conclude that the long-wavelength physics of 
FQHE is captured by introducing
familiar constitutive relations for the dielectric and diamagnetic response of
the Hall layer, but with conventional continuity conditions at the top and bottom
surfaces interchanged between perpendicular and
parallel components  of the electromagnetic fields.  

This work was supported in part by the National Science Foundation.  I thank
Jainendra Jain for discussions, and Steven Kivelson for emphasizing some time ago
that a conventional dielectric, diamagnetic medium could not reproduce 
any of the
characteristic phenomena associated with the fractional quantum
Hall effect.

\end{document}